\documentclass[reprint,aps,prx,amsmath,amssymb,amsfonts]{revtex4-2}
\usepackage[utf8]{inputenc}
\usepackage[T1]{fontenc}
\usepackage{bm}
\usepackage{graphicx}
\usepackage{dcolumn}% Align table columns on decimal point
\usepackage{tikz-cd}
\usepackage{xcolor}
\usepackage{hyperref} 
\usepackage{lineno}
\usepackage{soul}

\begin{document}
\preprint{APS/123-QED}
\title{Flux Jamming and Bimodal Dynamics in Bounded Spin Networks}

\author{Paula Mellado}
\email{paula.mellado@uai.cl}
\thanks{Corresponding author}
 \affiliation{Facultad de Ingeniería y Ciencias,\\ Universidad Adolfo Ibáñez, Santiago, Chile.}
\author{Hanu Arava}%
\affiliation{
Materials Science Division, Argonne National Laboratory, Lemont, Illinois, USA}%

%\date{\today}
%\linenumbers
\begin{abstract}
Finite-sized frustrated magnetic networks relax through pathways that are strongly shaped by their boundaries; yet, whether their kinetic barriers can be predicted from geometry alone has remained unclear. Here, we present a transfer matrix framework for finite artificial square ice that links local coordination to the effective energy landscape and the temperature-dependent barriers governing magnetic flux transport. By combining nonhomogeneous transfer matrices with the local coordination hierarchy, we identify the geometric bottlenecks that govern the relaxation dynamics, and show that the leading steady-state weight falls on charge-neutral, ice-rule configurations, which are thermodynamically preferred due to their reduced electrostatic energy and greater configurational multiplicity. The resulting two time-scale structure suggests bimodal behavior: (1) long intervals of trapping in charge-compensated manifolds are interrupted by (2) sudden avalanche-like relaxation events. Applied to Husimi-tree representations, the framework yields a low-temperature initial arrest probability that collapses across  forty generations onto a common scaling function of $g-\xi(T)$, with a crossover length $\xi \sim T^{-\nu}$ and $\nu=1$. These results show that boundary geometry alone can produce kinetic arrest and suggest intermittent transport, even in the absence of quenched disorder, reminiscent of athermal granular jamming. More broadly, our framework connects the coordination landscape, equilibrium energetics, and nonequilibrium relaxation in finite frustrated magnets and provides a route to understanding boundary-controlled dynamics in artificial spin ice.
\end{abstract}
\maketitle
\section{Introduction}
For many years, geometrically frustrated systems have been key to studying cooperative behavior in condensed matter physics \cite{toulouse2008frustration}. Early work on macroscopic degeneracy and residual entropy led to the water ice rules \cite{pauling1935structure} and, later, to their magnetic analogs, pyrochlore spin ice materials \cite{harris1997geometrical,ramirez1999zero, bramwell2001spin}. A major advance was the development of lithographically patterned macrospins, or Artificial Spin Ices, ASIs, \cite{wang2006artificial, nisoli2013colloquium, skjaervo2020advances}. By arranging single domain ferromagnetic nanomagnets into two dimensional lattices, ASIs not only replicate the local ice rule behavior found in bulk spin ices, but also enable direct, real space visualization of emergent charge defects, magnetic quasiparticles often referred to as magnetic monopoles \cite{castelnovo2008magnetic, ladak2010direct, mengotti2011real}. Although most prior studies have concentrated on bulk properties, practical devices—from neuromorphic computing architectures to nanomagnetic logic elements—are inherently finite, so their boundaries play a pivotal role in determining their large scale behavior.
At the same time, research on systems far from equilibrium has established strong conceptual links to the physics of disordered and amorphous granular materials. A key early advance in this area was the jamming paradigm introduced by Liu and Nagel \cite{liu1998jamming,cates1998jamming,liu2010jamming}, which unified the glass transition in supercooled liquids, the shear response of foams, and the mechanical confinement of granular particles. At the heart of athermal granular jamming lies the importance of the coordination landscape: when the average number of mechanical contacts per grain surpasses a critical geometric threshold, the system experiences an abrupt, disorder induced kinetic arrest, shifting from a flowing state to a rigid, immobilized configuration. A comparable coordination induced jamming has been demonstrated in artificial honeycomb spin ice composed of connected islands, where the focus moves from mechanical contacts per grain to magnetic constraints per vertex \cite{mellado2010dynamics}. To reveal how such a system traverses a complex, tightly constrained configuration space during transient relaxation, one must go beyond thermal activation and examine how the structural coordination restricts and ultimately suppresses transport dynamics \cite{udagawa2016out}.
\\
To describe how geometry and kinetics influence one another, research is turning to methods from algebraic graph theory \cite{schnakenberg1976network, felderhof1971spin}. In the case of magnets, graph theory offers a language in which magnetic junctions correspond to vertices and spins to edges \cite{chung1997spectral}. This framework has been successfully employed by several authors in ASI to predict ground-state degeneracy and zero-point entropy via graph permanents \cite{caravelli2021degeneracy}, quantify vertex frustration and defect creation in non-regular geometries \cite{morrison2013unhappy}, identify topologically protected kinetic manifolds \cite{lao2018classical}, and embed functional logic gates into nanomagnetic networks \cite{caravelli2020logical}. In this graph approach, the vertices represent the physical junctions or nodes of the lattice, while the edges represent the individual nanomagnetic spins \cite{caravelli2021degeneracy,estrada2013graph,perrin2016extensive}.  Yet most current graph-theoretic methods are largely static or rely on uniform periodic lattices, creating a key gap in extracting non-equilibrium kinetic relaxation from the spatially dependent coordination hierarchy set by physical boundaries.
\\
 Conventional statistical-mechanics treatments of artificial spin ice frequently assume translational symmetry; however, this premise fails for finite small clusters. Because truncated edges produce a highly nonuniform coordination landscape, boundary effects can strongly influence energy minimization. In hierarchical structures such as Cayley trees, Bethe lattices, and Husimi trees built from closed loops \cite{monroe2002blume}, exact recursive expressions for partition functions using transfer matrices \cite{baxter1985exactly} can be derived. This framework enables a direct examination of how discrete local constraints govern emergent macroscopic transport \cite{liebmann1986statistical, henley2010coulomb}. Experiments on finite nanomagnetic units accordingly show relaxation trajectories that depend sensitively on geometry and boundary conditions. Previous studies have demonstrated that nanomagnet arrays arranged on a square lattice and consisting of loops and vertices can operate as computational logic gates, where their function relies on guided thermal relaxation through specified sequences of vertices \cite{arava2018logic,saccone2023complex,saccone2023vertices,caravelli2020logical,gartside2022reconfigurable}. Later work showed that these relaxation pathways can be tailored by adjusting the boundary conditions of the constituent building blocks \cite{arava2019relaxation}. Direct observations of emergent monopole currents in finite arrays \cite{arava2020monpole}, together with measurements of energy relaxation pathways in finite square ice systems of different lateral dimensions \cite{arava2025energy}, have established that kinetic arrest is highly sensitive to both system size and boundary conditions \cite{hallen2022dynamical,libal2020quenched,saccone2022direct}. Comparable geometric tuning of magnetic relaxation has also been realized in kagome spin ice logic devices \cite{gypens2018balance}. Together with numerical reports of telegraph noise and avalanche-like relaxation \cite{mellado2010dynamics, shen2012dynamics, cumings2014focus, saccone2023real}, this highlights a theory–experiment gap: most models treat ASI as infinite and periodic, whereas devices are finite. Boundaries break translational symmetry and produce a position-dependent coordination that controls charge nucleation, transport, and annihilation. Closing this gap requires an analytically tractable finite-size framework that derives non-equilibrium barriers and kinetic arrest directly from the boundary-imposed coordination landscape.
\\
To bridge this gap and elucidate how boundary constraints shape non-equilibrium dynamics, this work presents an analytical framework that traces magnetic relaxation directly to the spatially varying coordination hierarchy ($z = 1, 2, 3, 4$) imposed by physical boundaries. Extending this construction beyond isolated building blocks to multi-generation hierarchical networks (the half-Husimi cactus), we present a non-homogeneous transfer-matrix formalism to deliver three key predictive advances. First, it derives how the boundary–bulk interface functions as a kinetic bottleneck, showing that rapid charge annihilation at $z \le 3$ boundaries coexists with prolonged trapping at $z = 4$ interior vertices, which is expected to generate bimodal kinetics and telegraph noise. Second, it shows that the principal eigenvector of the composite transfer operator is dominated by charge-neutral, ice-rule configurations, which minimize the free energy through their lower self- and Coulomb energy and their larger multiplicity W. For the Loop building block, where the basis resolves flux-closure states, this selects the chiral ground-state manifold observed experimentally. Finally, via finite-size scaling, we uncover a data collapse, demonstrating that boundary truncation alone—without extrinsic structural disorder—is sufficient to induce a jamming like crossover.

The paper is organized as follows. Section ~\ref{model} introduces the energetics of finite square clusters.  Section ~\ref{entropic} analyzes entropic selection in the highly degenerate manifold of chiral antiferromagnetic ground states. Section ~\ref{non} presents the algebraic construction of the non homogeneous transfer matrix chain. Section ~\ref{ground} details the ground state selection mechanism, and Section ~\ref{probabilistic} derives the probability of energy barriers in clusters. Section ~\ref{loop} applies the formalism to loop and vertex clusters, while Section ~\ref{large} extends it to large structures via the Husimi tree. Section ~\ref{finite} analyzes the finite-size scaling of the initial arrest probability, showing data collapse and the crossover characteristics of flux jamming in finite square ice clusters. Section ~\ref{conclusions} concludes and outlines future directions.
\section{\label{model}Finite square clusters}
In ASI, under the dumbbell approximation, a magnetic monopole is defined as the net sum of the magnetic charges $q^{(v)}=\pm 1$ at the ends of the magnets meeting at a vertex $v$ when that vertex violates the ice rules \cite{skjaervo2020advances}. In square ice, where bulk vertices have a coordination number of \(z = 4\), the lowest energy monopole carries a charge of \(Q_v=\sum_{i=1}^4 q_i^{(v)}=\pm 2\) (while monopoles with a charge of \(\pm 4\) are energetically prohibitive). In kagome or honeycomb spin ice, where \(z = 3\), a magnetic monopole has charge \(Q=\pm 3\) \cite{skjaervo2020advances}. 
\\
In a  finite square cluster, the journey of a magnetic monopole occurs across varying coordination environments, z, and the edges play a key role. 
Using a multipolar expansion of the magnetostatic interaction \cite{arava2026geometry,moller2009magnetic}, the energy of the system can be approximated by a vertex based Hamiltonian
\begin{equation}
E = \sum_{v} \frac{1}{2} \chi Q_v^2 + \sum_{u > w} \frac{Q_u Q_w}{r(u,w)}
\label{eq1}
\end{equation}
In  Eq.\ref{eq1}, the first term is the self energy of vertex $v$, with a total magnetic charge $Q_v$ where $\chi$ is the self energy constant. The second corresponds to the Coulomb interaction between magnetic charges at vertices $u$ and $w$, separated by a distance $r(u,w)$ (Madelung sum \cite{fumi1960extension}). Henceforth, energies will be expressed in units of the nearest-neighbor Coulomb interaction, and we set $k_B = 1$. In a magnetized magnetic cluster composed of a finite number \(n\) of connected magnets arranged on a square lattice, the range of possible vertex coordinations increases. Vertices with coordination \(z=3\) and dangling magnet vertices with \(z=1\) possess an intrinsic frustration of \(Q_v = \pm 1\), whereas \(z=2\) vertices can host higher energy charges of \(Q_v = \pm 2\). These boundary charges generate a magnetostatic potential that applies pressure on the \(z=4\) bulk region of the cluster as it evolves toward a minimum energy magnetic state.
\begin{figure}
\centering
\includegraphics[width=\linewidth]{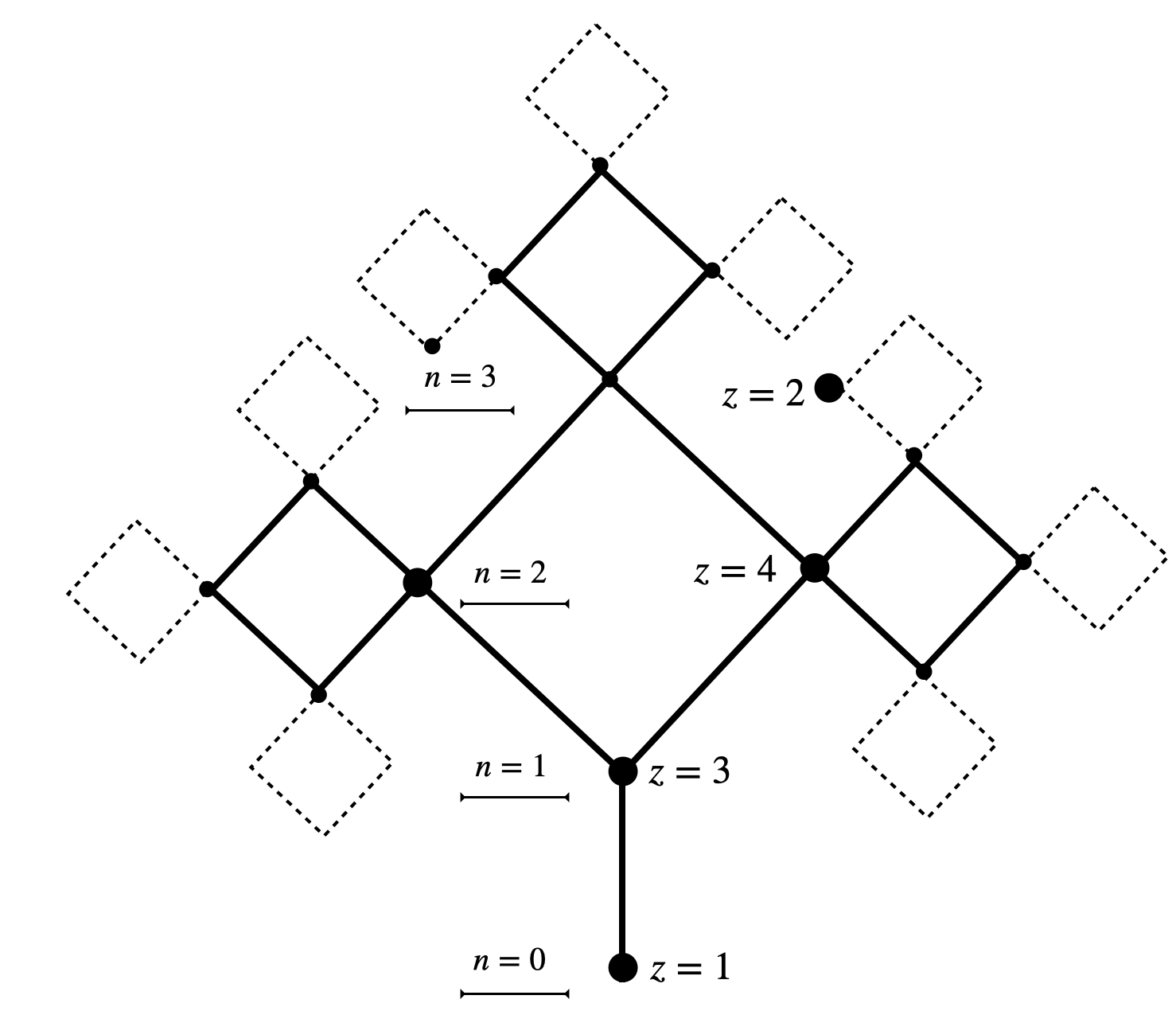}
\caption{Schematic representation of the half Husimi cactus geometry. The hierarchical tree is structured into a cascade of shared square plaquettes terminating at an open boundary. Internal bulk hubs feature a coordination number $z = 4$, while lateral nodes have $z = 2$. Each progressive generation introduces a shell with a path multiplicity that scales exponentially as $\omega_n = 2^n$.}
    \label{f1}
\end{figure}
\begin{figure}[htbp]
\centering
\vbox{
\hbox to \linewidth{\textbf{\textsf{a}}}
\vspace{2pt}
\hbox{\includegraphics[width=\linewidth]{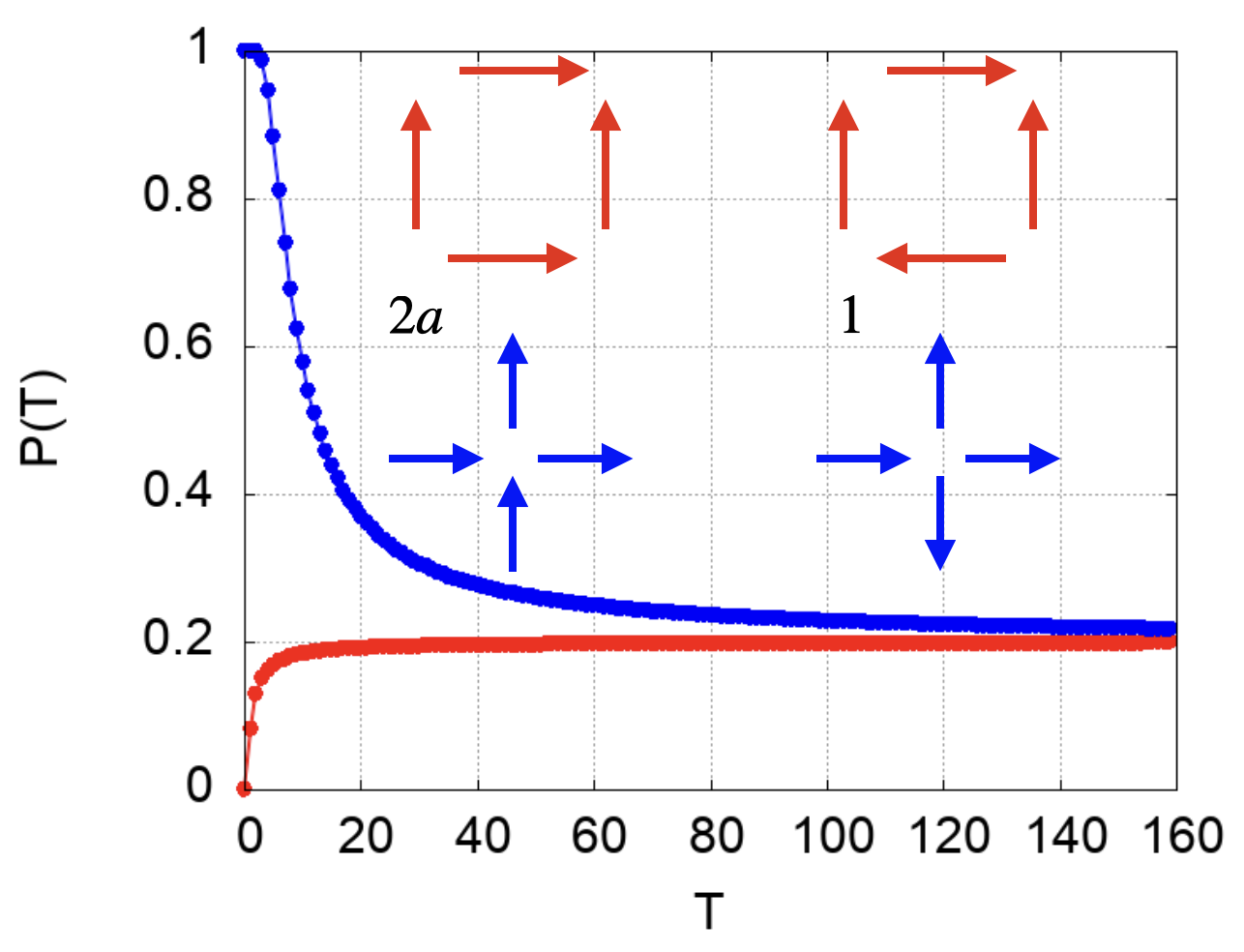}}}
\vspace{0.1cm} 
\vbox{
\hbox to \linewidth{\textbf{\textsf{b}}}
\vspace{2pt}
\hbox{\includegraphics[width=\linewidth]{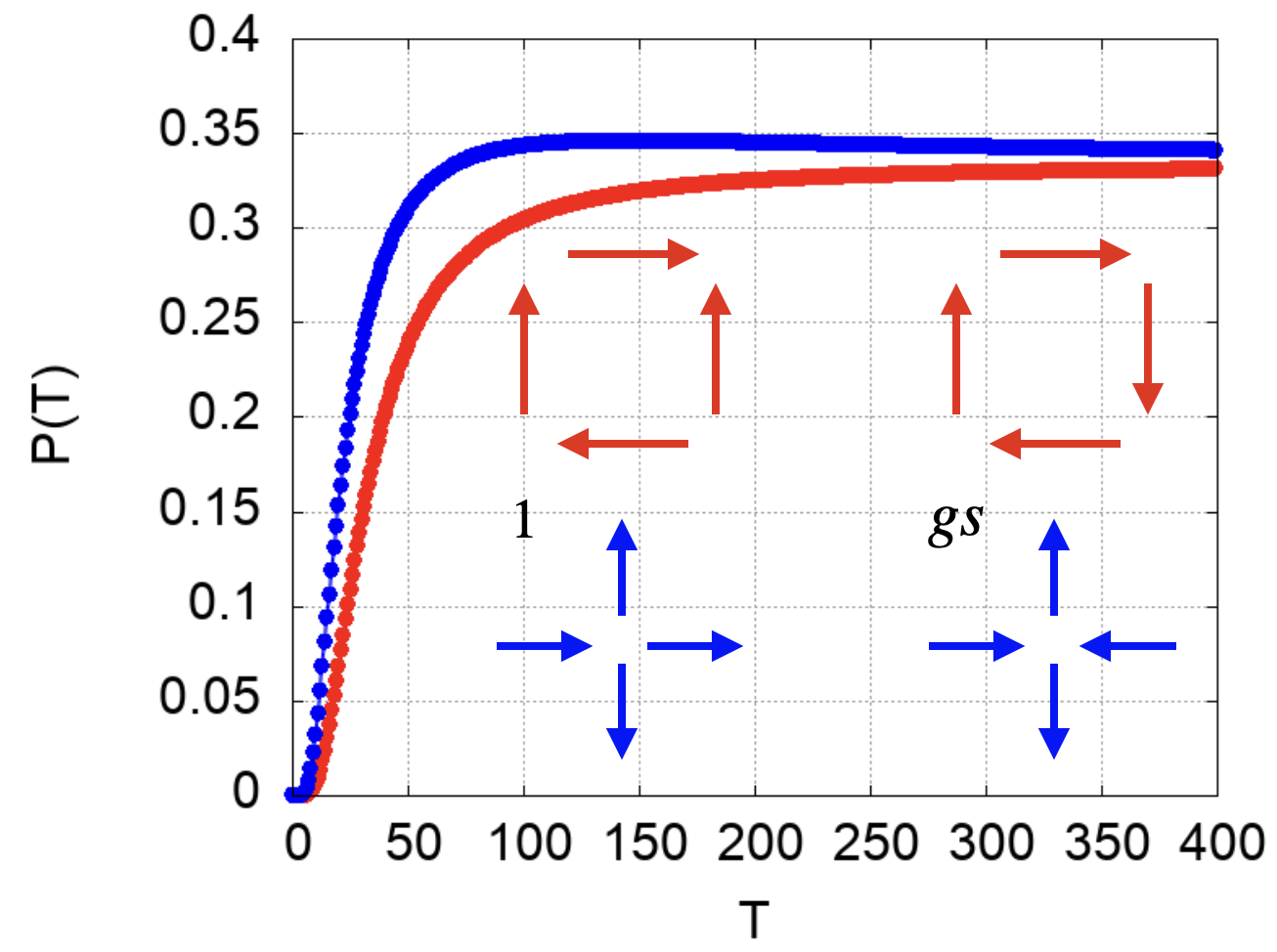}}
    }
\vspace{0.1cm}
\vbox{
\hbox to \linewidth{\textbf{\textsf{c}}}
\vspace{2pt}
\hbox{\includegraphics[width=\linewidth]{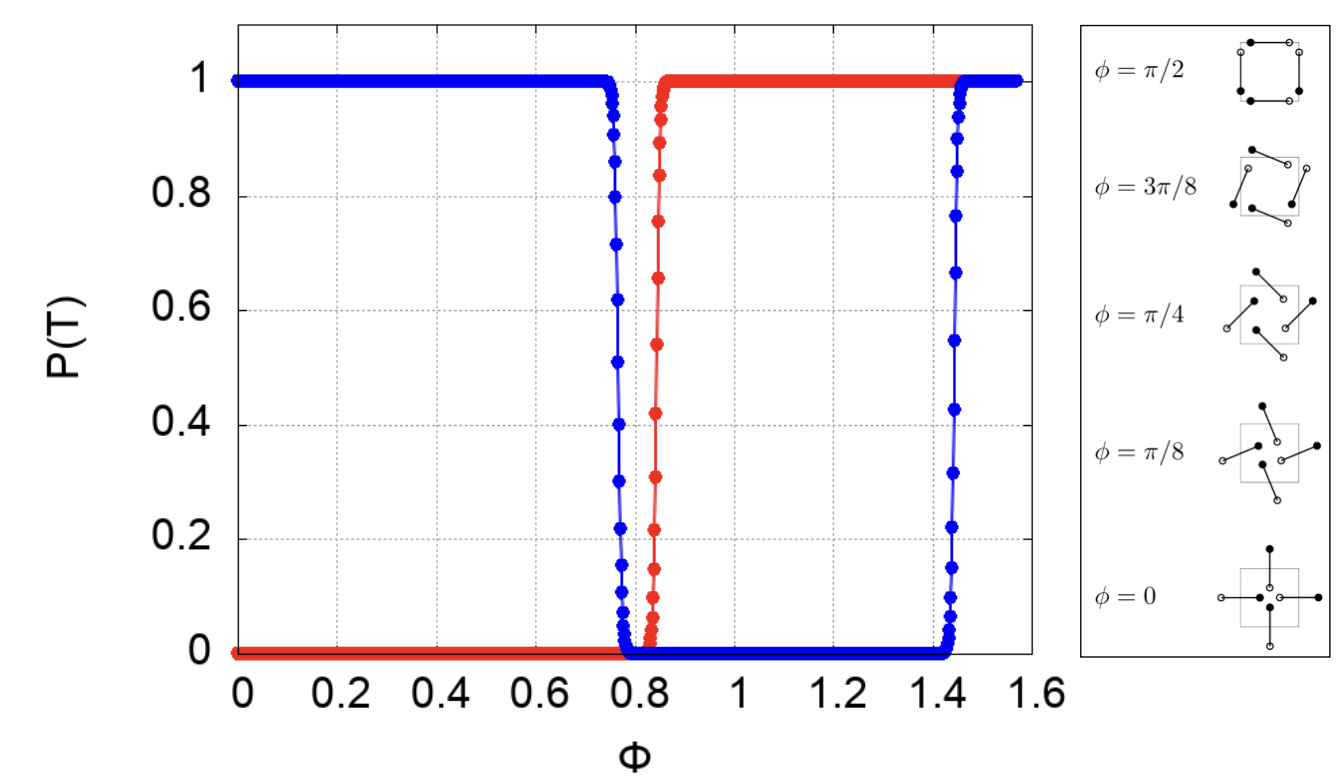}}
    }
\vspace{0.1cm}
\caption{Probability of energy barrier $P(T)$ as a function of temperature T ($k_B=1$),
\textbf{a.} in the transition $2a\to 1$ (Loop: red, Vertex: blue). Inset shows the loop and the vertex in one of the $2a$ and $1$ states. \textbf{b.} in the transition $1\to gs$ (Loop: red, Vertex: blue). Inset shows the loop and the vertex in one of the $1$ and $gs$ states. \textbf{c.} Probability of energy barrier as a function of $\phi$ at low temperatures ($T=10^{-3}$); the crossover is apparent at $\phi=\pi/4$ (Loop: red, Vertex: blue). Inset shows the transformation of a loop into a vertex as $\phi$ is tuned \cite{arava2026geometry}.}
\label{f2}
\end{figure}
\begin{figure}[htbp]
\centering
\vbox{
\hbox to \linewidth{\textbf{\textsf{a}}}
\vspace{2pt}
\hbox{\includegraphics[width=\linewidth]{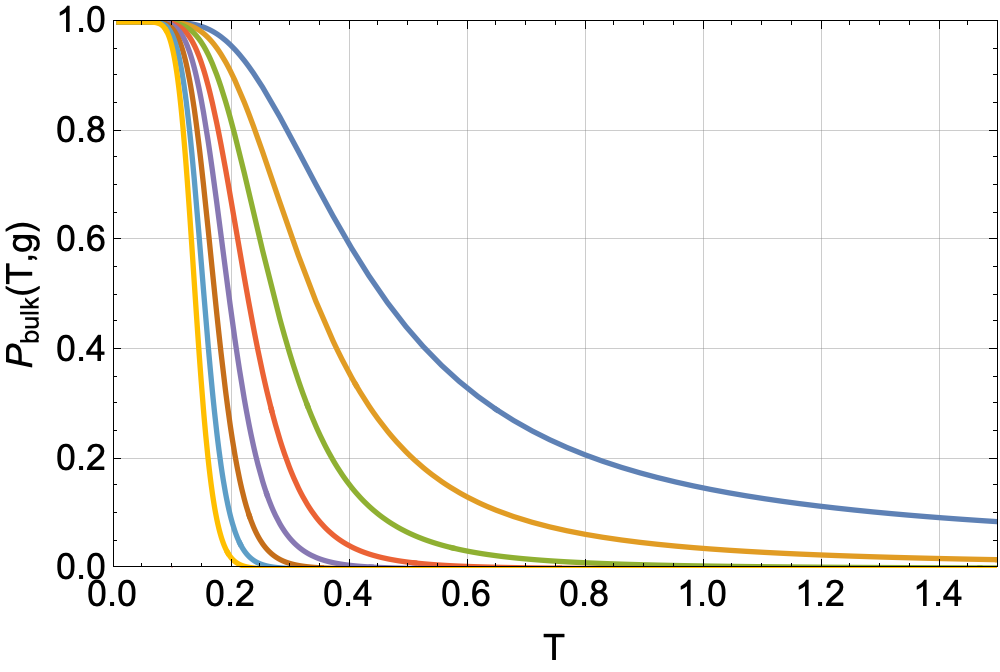}}}
\vspace{0.1cm} 
\vbox{
\hbox to \linewidth{\textbf{\textsf{b}}}
\vspace{2pt}
\hbox{\includegraphics[width=\linewidth]{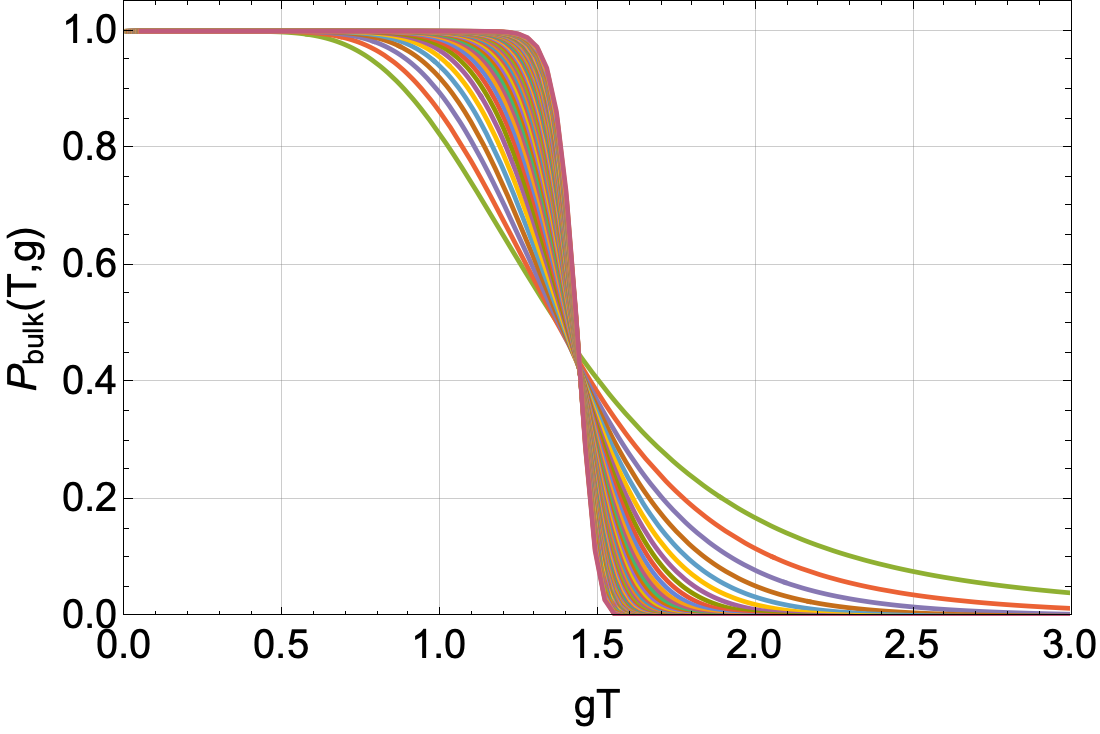}}}
\vspace{0.1cm}
\vbox{
\hbox to \linewidth{\textbf{\textsf{c}}}
\vspace{2pt}
\hbox{\includegraphics[width=\linewidth]{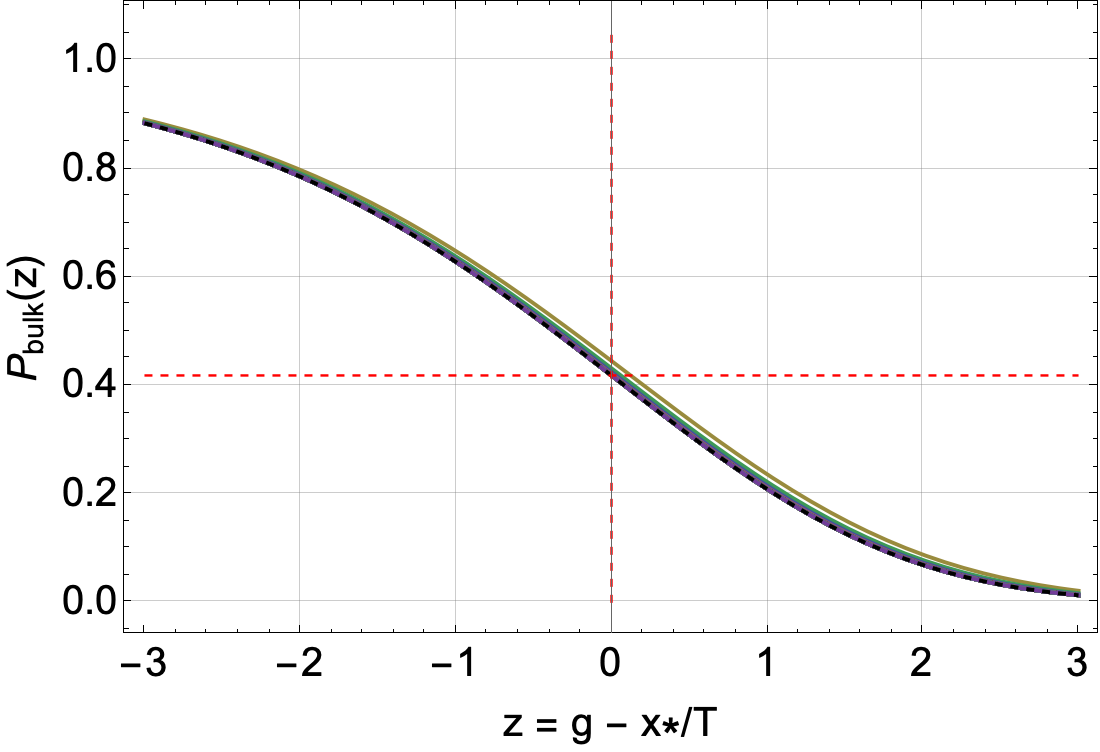}}}
\vspace{0.1cm}
\caption{Finite-size scaling of the interior bulk blocking probability
$P_{\rm bulk}(T,g)$ of the half-Husimi cactus, Eq.~(22), for $\chi=1.25$
($\Delta E_{\rm bulk}=4(\chi-1)=1$) and $k_B=1$. \textbf{a} $P_{\rm bulk}$ versus $T$ for $g=3,\dots,10$; the profile steepens and
moves to lower $T$ as $g$ grows. \textbf{b.} The same quantity for $g=5$--$50$ versus $x=gT$: all curves cross at
$x^*=4(\chi-1)/\ln 2\simeq1.44$, at the common height
$\prod_{m\ge1}(1+2^{-m})^{-1}\simeq0.419$, and sharpen around it with increasing $g$. \textbf{c.} Data collapse for $g=10$--$50$ versus $z=g-x^*/T$.
Dashed lines mark $z=0$ and $P_{\rm bulk}=0.419$. The collapsed curve approaches
$\mathcal{F}(z)$. $\Delta E_n = \Delta E_{\rm bulk}$ for all layers.}
\label{f3}
\end{figure}
\section{\label{entropic}Entropic Selection}
In a finite system, entropy $S$ signifies configurational degeneracy within a discrete state space \cite{greiner2012thermodynamics}. The entropic drive is the statistical weight of different energy manifolds and is defined by the number of microstates $W$ available at a specific energy $E$, $$S(E) = k_B \ln(W(E))$$ \cite{greiner2012thermodynamics}. In magnetized clusters, there are only a limited number of spin configurations that preserve maximum magnetization, so $W$ is small. The system will therefore relax into a state with residual entropy that minimizes the Helmholtz free energy, $F = E - TS$. 
\\
We have previously shown \cite{arava2025energy} that a range of small magnetic clusters built from square motifs, with sizes \(n \in (3,11)\), relax from an initially magnetized state into ground states (gs) that are characteristic of a chiral antiferromagnet (AFM). In these ground states, the bulk vertices organize into flux closure pots or loop shaped, flux closed configurations. Consequently, whenever possible, the system reduces its energy by aligning the spins in either a clockwise (CW) or counterclockwise (CCW) direction. Adjacent loops or plaquettes adopt opposite chiralities in order to reduce the overall Coulomb repulsion throughout the structure. This ground state forms the most highly degenerate set of configurations that both obey the ice rules and minimize dipolar interactions.
\\
As the clusters relax after being initialized in a non-equilibrium state by a magnetic field aligned at $\pi/4$ to the $x$-axis, the local transition rate between two microscopic spin configurations obeys a thermal Arrhenius form, \begin{equation} \Gamma_{i \to j} = \omega_{ij} \nu_0 \exp\left(-\frac{\Delta E_{ij}}{k_B T}\right) \cite{snyder2001spin}, \end{equation} where $\nu_0 \sim 10^9\text{--}10^{10}\text{ s}^{-1}$ represents the single-domain nanomagnetic attempt frequency, $\Delta E_{ij}$ is the magnetostatic activation barrier, and the multiplicity $\omega_{ij}$ quantifies the configurational degeneracy of the escape pathways connecting state $i$ to state $j$. Under local balance, the ratio of forward to backward transition rates is governed by the change in local Helmholtz free energy $\Delta F_{ij} = \Delta E_{ij} - T \Delta S_{ij}$, yielding \begin{equation} \frac{\Gamma_{i \to j}}{\Gamma_{j \to i}} = \frac{\omega_{ij}}{\omega_{ji}} \exp\left(-\frac{E_j - E_i}{k_B T}\right) = \exp\left(-\frac{\Delta F_{ij}}{k_B T}\right) \cite{jaubert2011magnetic}, \end{equation} where $\Delta S_{ij} = k_B \ln(\omega_{ij}/\omega_{ji})$ accounts for the entropic path asymmetry between configurations.  Counting each single-spin-flip link between the charge states $i$ and $j$ from either side, the multiplicities satisfy $W_i\,\omega_{ij}=W_j\,\omega_{ji}$, where $W_i$ is the number of microscopic spin configurations sharing the charge state $i$. Since forward and reverse flips cross the same saddle point, this gives $W_i\,\Gamma_{i\to j}=W_j\,\Gamma_{j\to i}$, so detailed balance holds identically with respect to the Boltzmann weight $W e^{-E/k_BT}$, and $\Delta S_{ij}=k_B\ln(W_j/W_i)$. An important point is that when mapping microscopic spin configurations onto vertex charge states, configurations sharing the same net charge may nonetheless allow different pathways to escape—an entropic degeneracy that is deliberately preserved in the calculation of the charge states’ effective trapping probabilities. In general, these elementary spin-flip rates define the rate generator of a continuous-time Markov Master Equation governing stochastic state evolution \cite{vankampen2007, glauber1963time}. Here, instead of calculating time-dependent trajectories, we focus on clarifying the mechanism behind arrested relaxation by computing the probability of local energy barriers, state survival probabilities, and bottlenecks arising from the hierarchical nature of spatial coordination. In magnetized square clusters, if the external field is applied at an angle of $\pi/4$ relative to the horizontal spin orientation, the two diagonal corner sites can become energetically degenerate. In this situation, the degeneracy factor is $\omega_{ij}=2$ enhancing the corresponding transition rate, thereby making the relaxation process approximately twice as likely to initiate via this particular entropic channel. When a spin flips, a charge $Q_v = \pm 2$ is created. While the energy $E$ increases, the number of ways to have a defect is much higher than the number of ways to be perfectly magnetized. This entropic gain helps the system overcome the initial energy barrier. 
\\
Overall, in a small cluster, the particular spatial arrangement of nodes produces a distinctive fingerprint for both the energy landscape and the entropic pathways, which is reflected in  (1) A   bias in the initial state that arises because the number of initial charges in the magnetized configuration depends on the cluster geometry. For example, a $3 \times 2$ spin cluster (three spins along the horizontal direction and two along the vertical) will have a different ratio of $z=3$ edge reservoirs to $z=4$ hubs than a $2 \times 2$ square. If the geometry yields a magnetized state in which the corner vertices carry $Q_v = \pm 2$, the driving force is much stronger than in a configuration where the corners start with $Q_v = 0$. In (2), the discreteness of $W$, since in a finite cluster, the density of states appears as a set of discrete peaks.
\section{\label{non}Non Homogeneous Transfer Matrix}
The relaxation dynamics of small square clusters can be rationalized using a position dependent transfer matrix \cite{baxter1985exactly,liebmann1986statistical} $\mathbf{T}_n(z)$ that connects the charge state of generation $n$, defined as the discrete structural shell of square plaquettes located at a geodesic distance of $n$ steps from the root vertex of the tree (open, dangling boundary link), with that of generation $n+1$
\begin{equation}
    \mathbf{T}_{n+1 \gets n} = \sum_{\{\sigma\} \in \text{gen n}, n+1} e^{-\beta E(\{\sigma\})} \lvert Q_{n+1} \rangle \langle Q_n \rvert
\end{equation}
where $\lvert Q_{n+1} \rangle$ denotes the column vector of charges in generation $n+1$, $\langle Q_n \rvert$ the row vector of charges in generation $n$, $\beta=\frac{1}{k_BT}$, T denotes temperature and the sum  runs over all microscopic spin configurations $\boldsymbol{\sigma}$ that map the charge state $\langle Q_n|$ to $|Q_{n+1}\rangle$. If there are $W$ different spin microstates with identical energy $E$ that all lead to the same charge transition $\langle Q_n| \to |Q_{n+1}\rangle$, then their individual Boltzmann factors add together, giving a multiplicative factor $W \cdot e^{-\beta \Delta E}$ for that matrix element. The factor $W$ represents the entropic contribution to the charge transition, with $S = k_B \ln W$. Finally, $E(\{\sigma\})$ is computed using \ref{eq1}. The matrix $\mathbf{T}$ operates on a charge basis $\mathcal{B} = \{0, \pm 1, \pm 2\}$. The elements $T_{ij}$ are Boltzmann weights $e^{-\beta \Delta E}$ for the transitions between states i and j. The $z=4$ bulk corresponds to a high energy barrier, where the transition $Q=0 \to \pm 2$ incurs an energy cost of approximately $\Delta E_{z4} \sim 4(\chi-1)$. In contrast, the $z=3$ nodes enable relaxation by acting as intermediate sites with a smaller barrier, $\Delta E_{z3} \sim 2(\chi-1)$ (see supplemental material for details \cite{supp}). The ratio of the associated matrix elements, $T_{z3}/T_{z4}$, determines the monopole conductance along the edges, indicating that relaxation preferentially proceeds through the $z=3$ boundary. The ice-rule (charge-neutral) manifold corresponds to the alignment of the state vector with the principal eigenvector of the composite operator $\prod \mathbf{T}_{n+1 \gets n}$ (Section V); the chiral ground state within it is resolved in Sec. VII.
\\
The geometry dependent properties of finite clusters can be systematically represented using a Husimi cactus built from square plaquettes \cite{monroe2002blume}. In the half Husimi cactus, the coordination hierarchy breaks translational symmetry. The state space is determined by the number of distinct charge configurations allowed at each vertex, which in turn is specified by z.  In a branching square cactus, internal sites have a coordination number $z=4$ (each connecting two squares), whereas boundary sites have $z<4$, mirroring the coordination environment of square clusters. In the half Husimi cactus shown in Fig.\ref{f1}, the multiplicity of available pathways attains its maximum at the $z = 3$ vertices, thereby enabling charge carriers to undergo local rearrangements until a bulk reconfiguration becomes energetically favorable.  By incorporating vertices with $z=1$, we capture the uncompensated spins located at the edges of finite networks. These terminal leaf nodes correspond to single spins with one end left dangling, not connected to any other junction. Such $z=1$ sites are the most unstable when subjected to an external magnetic field oriented at $\pi/4$ with respect to the horizontal spins.
\\
The transfer matrix $\mathbf{T}_{j \gets i}$ maps the state space of the vertex with coordination $z=i$ to the vertex with $z=j$. If vertex $i$ has $m$ possible charge states and vertex $j$ has $n$ states, $\mathbf{T}$ is an $n \times m$ rectangular operator (as shown in Table \ref{t1}) that facilitates the   flow of magnetic charges.   From the leaf toward the center of the tree, the coordination increases, and the matrix grows in dimension because there are more ways to arrange spins to create different charge states. This represents the reservoir capacity of the edges. As the system enters the $z=4$ bulk, the manifold constricts again to only the even charges $\{0, \pm 2\}$. This serves as a   bottleneck. The $3 \times 4$ matrix eliminates the frustrated edge states with $Q=\pm 1$, forcing the system to obey the ice rules ($Q=0$). The full relaxation process is represented as a sequence of operators that begins at the $z=1$ dangling root link. The initially magnetized state is carried into the bulk via a succession of rectangular transfer operators (see Table \ref{t1}).
\\
To analyze the relaxation spectrum, the dimensional mismatch between the incoming boundary states and the bulk manifolds must be reconciled. We track this dimensionality explicitly by mapping the charge manifolds across successive coordination shells. The root generation ($n=0$) at the boundary leaves ($z=1$) is defined by two possible   charges ($\pm 1$), spanning a $2$ dimensional state space. Moving inward, the $z=3$ intermediate nodes encompass four charge states ($\pm 1, \pm 3$), spanning a $4$ dimensional space. The subsequent $z=4$ junctions  contract to a $3$ dimensional manifold supporting charges ($0, \pm 2$). Finally, at the $z=2$ edge, the low energy subspace has three charge configurations ($0, \pm 2$), defining a $3$ dimensional destination space. Consequently, the sequential forward mapping from the $z=1$ leaves to the boundary $z=2$ for the Husimi cactus of Fig.\ref{f1} is rectangular, constructed via the product of these coordination shifting tensors as $\mathcal{M} = \mathbf{T}_{2\leftarrow 4}\mathbf{T}_{4\leftarrow 3}\mathbf{T}_{3\leftarrow 1}$. Because $\mathcal{M}$ maps a $2$-dimensional initial boundary vector to a $3$-dimensional edge state, it forms a $3 \times 2$ matrix. To close the dynamic loop, we introduce a rectangular $2 \times 3$ boundary closure operator 
\begin{equation} \mathcal{P}_{\text{cl}} = \sigma_x \mathcal{M}^{\mathsf{T}}, 
\end{equation} where $\mathcal{M}^{\mathsf{T}}$ denotes the $2 \times 3$ transposed map that projects the low-energy bulk states back onto the $z=1$ boundary leaves, and the first Pauli matrix $\sigma_x$ serves as the boundary swap operator. Pre-multiplying by $\sigma_x$ incorporates the reflection-induced reversal of magnetic moments by explicitly interchanging the boundary charge labels, thereby enforcing a chirality exchange $(+1) \leftrightarrow (-1)$ when exiting the boundary \cite{supp, berman1994nonnegative}. Consequently, the full composite transfer matrix is defined as the square operator
\begin{equation}\tilde{\mathcal{M}} = \mathcal{P}_{\text{cl}} , \mathbf{T}_{2\leftarrow 4}\mathbf{T}_{4\leftarrow3}\mathbf{T}_{3\leftarrow1} = \sigma_x \left( \mathcal{M}^{\mathsf{T}} \mathcal{M} \right)\end{equation}
reducing the global network dynamics to a square boundary to boundary transfer matrix. Since the matrix elements  are each positive sums of Boltzmann weights, all entries of $\tilde{\mathcal{M}}$ are strictly positive. As a result, $\tilde{\mathcal{M}}$ falls under the conditions of the Perron–Frobenius theorem, which ensures a single real dominant eigenvalue and an associated steady-state leading eigenvector that determines the prevailing macroscopic charge distribution \cite{supp, berman1994nonnegative}.
\begin{table*}[htbp]
\centering
\caption{Non homogeneous transfer matrices across the coordination hierarchy of the half Husimi cactus.}
\label{t1}
\small
\setlength{\tabcolsep}{10pt}
\renewcommand{\arraystretch}{1.3}
\begin{tabular}{lccc}
\hline \hline
Transition Type & Dimension ($n \times m$) & Charge Basis Mapping & Physical Role \\ \hline
$z=1 \to z=3$ & $4 \times 2$ & $\{\pm 1\} \to \{\pm 1, \pm 3\}$ & Boundary Flux Initiation \\
$z=3 \to z=4$ & $3 \times 4$ & $\{\pm 1, \pm 3\} \to \{0, \pm 2\}$ & Forward   Bottleneck \\
$z=4 \to z=4$ & $3 \times 3$ & $\{0, \pm 2\} \to \{0, \pm 2\}$ & Core Bulk Relaxation \\ 
$z=4 \to z=2$ & $3 \times 3$ & $\{0, \pm 2\} \to \{0, \pm 2\}$ & Reverse Lateral Projection \\ \hline \hline
\end{tabular}
\end{table*}
\section{\label{ground}Ground State Selection}
The evolution of a magnetized half Husimi cactus is determined by the spectral characteristics of its composite operator \cite{eggarter1974cayley}. The steady state charge distribution converges to the principal eigenvector of $\tilde{\mathcal{M}}$,  $\vec{v}_1$ associated with its maximum eigenvalue $\lambda_{max}$ \cite{chung1997spectral,berman1994nonnegative}.
The Q=0 weight of $\mathcal{M}\vec{v}_1$ indicates selection of the ice-rule manifold by the free energy: charge-neutral configurations carry no vertex self-energy or Coulomb cost and, in addition, have a larger multiplicity than monopole-bearing ones, which enhances their statistical weight. Because the charge basis does not distinguish among ice rule states, the chirality of the ground state is resolved only in the Loop example of Sec.VII. To show that the principal eigenvector of $\tilde{\mathcal{M}}$ is dominated by the ice rule configuration, we must analyze the exact algebraic structure of transfer matrices. Every element within $\mathbf{T}$ is proportional to a local Boltzmann weight $e^{-\beta \Delta E}$. Consider $k\equiv e^{-4\beta(\chi - 1)}$, the weight of one spin flip creating a monopole–antimonopole pair in a neutral neighborhood, where $\chi$ is the vertex self energy. At low temperatures, charge neutral configurations have weight $\sim 1$, while frustrated monopole states ($Q=\pm 2$) are suppressed as $k^{1/2}$ or k (see Tables SI, SII and SIII in supplemental material \cite{supp}). The generation of a frustrated $z=3$ node configuration demands a local restructuring energy of $\Delta E_{3\gets1} = 4(\chi -1)$. Expressing this energy cost in terms of $k$, the Boltzmann weight evaluates to
\begin{equation}
    e^{-\beta \Delta E_{3\gets1}} =k
\end{equation}
Because the $z=1$ leaf space is $2$ dimensional ($\pm 1$) and the $z=3$ junction space is $4$ dimensional ($\pm 1, \pm 3$), $\mathbf{T}_{3 \gets 1}$ becomes a rectangular $4 \times 2$ matrix. Ordering the column basis as $(+1, -1)^T$ and the row basis as $(+1, -1, +3, -3)^T$ yields,
\begin{equation}
    \mathbf{T}_{3 \gets 1} = \begin{pmatrix} 
        1 & 0 \\ 
        0 & 1 \\ 
        k& 0 \\ 
        0 & k
    \end{pmatrix},
\end{equation}
where the unfrustrated pathways carry unsuppressed unit weights, while the highly frustrated triple charge rows are gated by the bottleneck $k$ (the explicit form of the other matrices is available in the  Supplemental material \cite{supp}).
When the transfer matrices are multiplied to construct $\mathcal{M}$, this results in a sum over products. The $Q=0$ paths consist of products of $1 \times 1 \times 1$, while any path leading to a $Q=\pm 2$ bulk state involves at least one factor of $k^{1/2}$. In the limit $\beta \to \infty$ every column of the resulting composite matrix $\mathcal{M}$ points almost entirely toward the $\vec{Q}=0$ vector.  If the middle row of $\mathcal{M}$ is significantly larger than the others, then for any input vector $\vec{x}$, the output $\mathcal{M}\vec{x}$ will be heavily biased toward the middle component. Since all elements of $\tilde{\mathcal{M}}$ are positive, there exists a unique largest eigenvalue whose eigenvector has strictly positive components \cite{chung1997spectral}. Because the energy sink of the ice rules provides the path of least resistance (highest weights), the principal eigenvector settles there. This means that as the system evolves, it doesn't matter how frustrated the $z=1$ boundary is because the $z=4$ bulk bottleneck will eventually drain all the $Q \neq 0$ components of the state vector. As a result, the $\vec{Q}=0$ state becomes the only surviving component of the state vector over time. 
\section{\label{probabilistic}Probabilistic Barrier Analysis}
The likelihood of overcoming an energy barrier is described by a stochastic transition matrix $\mathcal{P}$, with elements $\mathcal{P}_{ij} = T_{ij} / \sum_k T_{kj}$. In the $z=4$  junctions, for instance, the probability of a monopole producing transition is suppressed by the factor $\mathcal{P}(0 \to \pm 2) = \frac{k}{1 + 2k} \sim e^{-4\beta(\chi -1)}$ \cite{supp}. This exponential suppression is the cause of the observed kinetic arrest or intermittence in finite ASI clusters \cite{arava2025energy}. 
\\
For a given transition from coordination layer $i$ to layer $j$, the matrix element $(T_{j\leftarrow i})_{\gamma \alpha}$ represents the unnormalized Boltzmann weight connecting an initial state $\alpha \in \mathcal{B}_i$ to a subsequent state $\gamma \in \mathcal{B}_j$. The conditional probability $\mathcal{P}(\gamma_j \mid \alpha_i)$ that the system transits into state $\gamma$ given that it occupies state $\alpha$ is 
\begin{equation}
\label{eq:proba}
    \mathcal{P}(\gamma_j \mid \alpha_i) = \frac{(T_{j\leftarrow i})_{\gamma \alpha}}{\sum_{k} (T_{j\leftarrow i})_{k \alpha}},
\end{equation}
where the denominator sums over all allowed channels in the output manifold $\mathcal{B}_j$. To quantify the local kinetic stability during the cluster's relaxation, the probability of the system remaining in its current configuration $\alpha$ before advancing to the next coordination layer is defined via a localized survival rate. By choosing the baseline weight of the held state to be unity, the conditional probability of remaining in state $\alpha$, or equivalently, the probability associated with an energy barrier, can be expressed as
\begin{equation}
    \mathcal{P}(\text{stay} \mid \alpha) = \frac{1}{1 + \sum_{k} T_{k\alpha}},
    \label{eq:stay}
\end{equation}
where the sum in the denominator runs over all allowed transition channels $k$ within the subsequent rectangular transfer matrix layer, and $\sum_{k} T_{k\alpha}$ represents the cumulative escape weight out of the local manifold. If the system is sitting in state $\alpha$ and faces a large energy barrier to move to any subsequent state, the Boltzmann weights $T_{k\alpha} = e^{-\beta \Delta E_k}$ will be very small ($T_{k\alpha} \rightarrow 0$). The denominator reduces to $1$, making $\mathcal{P}(\text{stay}) \rightarrow 1$. The system is completely trapped or kinetically arrested.  
\\
The question of whether the system leaves or stays is contained within the columns of the rectangular matrix. Even if individual energy barriers are low, if there are very few allowed geometric paths, the sum remains small, capturing how low coordination nodes limit the relaxation rate. For instance, in the $z=4$ bulk matrix ($\mathbf{T}_{4 \gets 3}$), $\mathcal{P}(\text{stay in } Q=0) \sim \frac{1}{1 + 2k}$. Since $k=e^{-4\beta(\chi - 1)}$ is very small at low temperatures, the probability of staying in the neutral state is nearly 1, and the probability of jumping the barrier to create a $Q=\pm 2$ monopole is nearly 0. 
\\
Although individual spin flips follow standard Arrhenius activation, the global relaxation of finite networks is non Arrhenius, a hallmark of out of equilibrium dynamics \cite{bouchaud1998out}. Within the half Husimi cactus representation, this behavior is driven by the spectral properties of the non homogeneous transfer operator. Local coordination mismatches serve as   bottlenecks—appearing as high energy barriers that generate kinetic arrest and intermittent relaxation across multiple timescales. The overall relaxation rate is governed by how strongly the magnetized state vector $\lvert \psi_{\text{pol}} \rangle$ initially projects onto the $z=2$ boundary manifold. This gives rise to two possible scenarios.
In Scenario I, the external magnetic field prepares a frustrated, highly charged boundary state ($Q_{z2} = \pm 2$) situated at a local energy maximum on the potential energy surface. This configuration drives a trajectory where the high initial monopole chemical potential immediately dissipates into the adjacent $z=3$ reservoir. Consequently, the initialization barrier is bypassed, and the initial blocking probability vanishes. 
\begin{equation}
    \mathcal{P}(\rm stay)_{\text{init}} \sim 0 \quad (\text{Scenario I}),
\end{equation}
and the localized transfer weight approaches unity, triggering ballistic like monopole (monotonic \cite{arava2025energy}) propagation across the network boundaries.
\\
In Scenario II, the boundary is compensated ($Q_{z2} = 0$), initializing the system in a metastable neutral valley where relaxation requires the creation of a monopole antimonopole pair. The probability of initial kinetic arrest is determined by the weight of the neutral state relative to the total configurational weight 
\begin{equation}\mathcal{P}(\rm stay)_{\text{init}} = \frac{1}{1 + 2k} \quad (\text{Scenario II})\end{equation}
In the athermal limit ($T \to 0, k \to 0$), the initial arrest probability $\mathcal{P}(\rm stay)_{\text{init}} \to 1$ is expected to lead to intermittent relaxation governed by energetic barriers. 
\\
The combination of Scenario I and II dynamics is expected to yield a bimodal kinetic distribution. The network remains locked in prolonged kinetic arrest within the neutral manifold until a critical thermal fluctuation surmounts the governing activation barrier, triggering a rapid, avalanche like cascade that dynamically relaxes the system \cite{monthus1996models,sethna2001crackling,ritort2003glassy}
\subsection{Connecting discrete spatial transfer operators to continuous-time Markov kinetics}
\label{subsec:kinetic_generator_derivation}
To establish a continuous-time stochastic description from spatial transfer matrices, we construct a Markovian rate generator that governs non-equilibrium relaxation across the coordination hierarchy ($z = 1, 2, 3, 4$). Let $\mathbf{P}(t) = [P_1(t), P_2(t), \dots, P_N(t)]^T$ represent the time-dependent probability vector over the reduced charge state basis $\{Q_v\}$, where $N$ is the dimension of the state space, representing the total number of distinct charge states (or configurations) in the basis $\{Q_v\}$ being tracked. The continuous-time Markov Master Equation governing state evolution, $\frac{d P_i(t)}{dt} = \sum_{j \neq i} \left[ \Gamma_{i \leftarrow j} P_j(t) - \Gamma_{j \leftarrow i} P_i(t) \right]$, is expressed in compact matrix notation as \cite{vankampen2007, glauber1963time}: \begin{equation} \frac{d \mathbf{P}(t)}{dt} = \mathbf{W} \mathbf{P}(t), \end{equation} where $\mathbf{W}$ is the $N \times N$ kinetic rate generator. Normalizing the spatial transfer matrix elements $T_{i \leftarrow j}$ over all available escape channels yields the matrix representation of the conditional layer transition probabilities [Eq.~(\ref{eq:proba})]: \begin{equation} K_{ij} = \frac{T_{i \leftarrow j}}{\sum_k T_{k \leftarrow j}}. \label{eq:normalized_jump_matrix} \end{equation} Multiplying by the single-domain attempt frequency $\nu_0 $ sets the off-diagonal rates $W_{ij} = \Gamma_{i \leftarrow j} = \nu_0 K_{ij}$ ($i \neq j$). Probability conservation ($\sum_i W_{ij} = 0$) enforces diagonal entries $W_{jj} = -\nu_0 \sum_{i \neq j} K_{ij} = \nu_0 (K_{jj} - 1)$. Combining diagonal and off-diagonal entries via the Kronecker delta yields $W_{ij} = \nu_0 (K_{ij} - \delta_{ij})$, establishing the mapping: \begin{equation} \mathbf{W} = \nu_0 (\mathbf{K} - \mathbf{I}), \label{eq:operator_generator_relation} \end{equation} where $\mathbf{I}$ is the $N \times N$ identity matrix.This makes it possible to determine the lifetime $\tau_j$ of state $j$ (in seconds) as
\begin{equation}
\tau_j = \frac{1}{|W_{jj}|} = \frac{1}{\nu_0 (1 - K_{jj})}.
\label{eq:physical_lifetime}
\end{equation}
Equation~(\ref{eq:physical_lifetime}) shows that when energy barriers are large ($\Delta E \gg k_B T$) or the coordination is high ($z = 4$), escape processes are greatly suppressed ($K_{jj} \to 1$). Consequently, the residence time diverges ($\tau_j \to \infty$), providing the condition for kinetic arrest.

The relaxation spectrum follows from the eigenvalues $\mu_k$ of $\mathbf{W}$, related to those of $\mathbf{K}$ by $\mu_k=\nu_0(\lambda_k-1)$. Thus $\lambda_1=1$ gives $\mu_1=0$, yielding a stationary state with $\mathbf{W}\mathbf{P}_{\text{ss}}=0$. For $k\ge2$, the relaxation times are:
\begin{equation}
\tau_k = -\frac{1}{\Re(\mu_k)} = \frac{1}{\nu_0 (1 - \Re(\lambda_k))}.
\label{eq:relaxation_time_spectrum}
\end{equation}
where $\Re(\lambda_k)$ denotes the real part of the $k$-th eigenvalue $\lambda_k$ of the transition matrix $\mathbf{K}$. When the spectral gap is small, $\epsilon = 1 - \Re(\lambda_k) \ll 1$, using the linearization $1 - \lambda_k \approx -\ln(\lambda_k)$ yields the log-ratio relationship in the appropriate limit, thereby connecting discrete spatial transfer operators to continuous-time Markov kinetics.

Two descriptions of the kinetics are used. Eq.~(2) gives the microscopic single-flip rates; it fixes the time scales $\tau_{\rm fast}$ and $\tau_{\rm slow}$ and satisfies detailed balance (Eq.~3). The matrix $\mathbf K$ of Eq.~(14) is the column-normalized form of the same rates, used for the occupation and survival probabilities. The two coincide when the escape weights are small, $\sum_{i\neq j}T_{i\leftarrow j}\ll1$, which is the regime where arrest is defined, and they differ only near the crossover $2^n e^{-\Delta E/k_BT}\simeq1$, where the scaling variable, and hence $\xi(T)$, is the same. Detailed balance is therefore claimed for Eq.~(2), not for the normalized $\mathbf K$. The rectangular $\mathbf T_{j\leftarrow i}$ are the off-diagonal blocks of a square matrix whose diagonal blocks hold the stay probabilities; the $\lambda_k$ of Eq.~(17) are the eigenvalues of this matrix, and $\tilde{\mathcal M}$ is its reduction to the $z=1$ sector. Grouping microstates by their vertex charges is exact only if all microstates of a charge state have the same escape pathways; otherwise it is an approximation.
\section{\label{loop}The loop and the vertex building blocks}
Before extending the previous results to more complex architectures, we begin by applying it to the Loop and Vertex clusters that were recently characterized in experiments \cite{arava2026geometry}. In these clusters, four magnetic elements are placed at the midpoints of the sides of square plaquettes of side \(a\). Each nanomagnet has uniform magnetization, with its moment constrained to its long axis, forming an Ising like degree of freedom. The four Ising moments give $2^4 = 16$ possible magnetic configurations. The relaxation pathways of the Loop and Vertex clusters and their rotated versions were investigated in our earlier work \cite{arava2026geometry}. The magnetostatic interaction between the magnets is approximated using a multipolar expansion \cite{arava2026geometry}. Each magnet is oriented along the unit vector $(\cos{(\theta_m + \phi)}, \sin{(\theta_m + \phi)})$, where the fixed azimuthal angle $\theta_m=(m-1)\frac{\pi}{2}$, ($m=1,2,3,4$) represents the orientation of the center of the magnet with respect to the centroid of the cluster and $\phi \in [0, \pi/2]$ is the in plane rotation angle relative to its local $z$ axis at the magnet center. Vertex and loop geometries occur at $\phi = 0$ and $\phi = \pi/2$, respectively, as shown in Fig.\ref{f2}c. Their possible magnetic configurations fall into four inequivalent families with equal energy within each set, labeled $0$, $1$, $2a$, and $2b$ (supplemental figure 3 in \cite{arava2026geometry}). Initially, Loop and Vertex are magnetized within their plane at an angle of $\pi/4$ with respect to their horizontal spins. In both cases, the magnetized states belong to family 2a. 
\\
In the Loop, the state 2a  comprises four distinct magnetic configurations, each characterized by a pair of magnetic monopoles of opposite charges located on two opposite diagonals of the square. This state falls into Scenario I, anticipating a vanishing initial blocking probability. The number of charge states determines the number of columns of $\mathbf{T}_{1\leftarrow 2a}$, which describes the first stage of the relaxation dynamics. A single spin flip starting from any of the 2a states yields eight distinct resulting configurations, which together are designated as family 1; these outcomes define the number of rows of the Loop transfer matrix $\mathbf{T}_{1\leftarrow 2a}^\ell$ (see \cite{supp}) which has dimension $8\times 4$. In the next relaxation step, a further spin flip moves the system from a state in family 1 either to state 0—this corresponds to the Loop ground state family, which includes the two flux closed configurations (CW, CCW), or to family 2b, which contains the two higher energy magnetic configurations in the Loop featuring monopoles of identical sign located on opposite diagonals. Consequently, the rectangular matrix $\mathbf{T}_{gs\leftarrow 1}^\ell$ has dimensions $4\times 8$ (see \cite{supp} for explicit matrices).
\\
The magnetized state of the Vertex comprises the four neutral magnetic states 2a, as shown in \cite{arava2026geometry}. This situation falls into Scenario II. As in the Loop case, once a single spin flips, the system can evolve into eight distinct charge configurations: four of these have charge -1 at the central node, while the remaining four have charge +1. This stage gives rise to the $8\times 4$ matrix $\mathbf{T}_{1\leftarrow 2a}^v$. In the next relaxation stage, an additional spin flip drives the system from a state in family 1 to either of the two possible charged states 0, which are very high in energy (as they host charge $\pm 4$ at the central node), or to the ground state family 2b, which comprises the two neutral states of lowest energy. Overall, the rectangular matrix $\mathbf{T}_{gs\leftarrow 1}^v$ has dimensions $4\times 8$ (see \cite{supp}). 
\\
Figure~\ref{f2}a shows the probability of encountering an energy barrier when the Loop (red) and Vertex (blue) configurations relax from the magnetized initial state to the family of states $1$, as a function of temperature. In the high temperature limit, the barrier probability for both geometries is $\mathcal{P}(T)=\frac{1}{1+4e^0}=\frac{1}{5}$, where the factor of 4 arises from summing identical Boltzmann weights across the columns of $\mathbf{T}_{1\leftarrow 2a}^\ell$ and $T_{1\leftarrow 2a}^v$. As the temperature T lowers, the Loop and Vertex exhibit increasingly distinct behaviors: the barrier crossing probability for the Loop approaches zero, while for the Vertex it tends toward 1, characteristic of monotonic versus intermittent relaxation \cite{arava2026geometry}. In the subsequent stage of the evolution, from state 1 to the gs, corresponding to configurations $2a$ in the Loop and $2b$ in the Vertex, a single spin flip is energetically favorable in both structures. Consequently, at low temperatures, Figure~\ref{f2}b shows that both barrier probabilities approach $0$.
\\
Figure~\ref{f2}c shows how the barrier probability varies with the geometrical parameter $\phi$ as the Loop ($\phi=\pi/2$) and Vertex ($\phi=0$) configurations gradually transform into one another at low temperatures. One might expect $\mathcal{P}(stay)$ to depend strongly on the local orientation of the magnets. Instead, we find that varying $\phi$ alters the barrier probability only within a narrow range around $\phi=\pi/4$, where a geometrical crossover between monotonic and intermittent relaxation occurs \cite{arava2026geometry}.
\\
At low temperatures, the system's ground state is encoded in the $4\times 4$ composite transfer operator. For a Loop, the first two rows of $\mathcal{M}^\ell$ govern transitions from state 1 to the degenerate, charge neutral ground state manifold ($Q=0$). The elements of these rows increase as temperature decreases, whereas the bottom two rows—governing transitions to high energy transient states ($2b$) asymptotically vanish. The energy spectrum of $\mathcal{M}^\ell$ yields a dominant eigenvalue $\lambda_{\max}$ with a symmetric principal eigenvector $v_1^\ell=\frac{\sqrt{2}}{2}(1,1,0,0)$. This state represents the normalized probability amplitudes of the degenerate ground state manifold reached after relaxation from an initially magnetized, monopole bearing configuration to a divergence free, flux closure state. Because the underlying Hamiltonian preserves time reversal symmetry and contains strictly positive Boltzmann weights, the principal eigenvector partitions its weight equally between the clockwise and counterclockwise chiral configurations. The nonzero projections of $v_1^\ell$ map the asymptotic state onto the divergence free manifold, indicating that the initial monopole chemical potential has been entirely dissipated into circulating flux.
\\
For a Vertex, the principal eigenvector rotates in basis space to $v_1^v = \frac{\sqrt{2}}{2}\left(0, 0, 1, 1\right)$, where the non zero components correspond to two degenerate, charge neutral Type II spin ice configurations \cite{skjaervo2020advances} related by time reversal symmetry. Because this vertex is prepared in a microscopically compensated state, it encounters severe initial kinetic arrest; in the low temperature limit, the initial blocking probability saturates to unity. Consequently, the system exhibits prolonged residence times trapped within the initial magnetized valley due to the high activation energy required to generate the highly frustrated monopole intermediates (Type III and Type IV states \cite{skjaervo2020advances}) necessary for the transition. Only when a thermal fluctuation surmounts this defect creation barrier does the configuration undergo a rapid, avalanche like unwinding, dynamically restoring time reversal symmetry within the symmetric manifold defined by $v_1^v$.
\\
Ultimately, this transfer matrix formalism captures both annihilation driven and creation governed relaxation trajectories. The principal eigenvector demonstrates that regardless of initial magnetization, long term relaxation invariably drives the system toward a unique destination. Once local energy barriers are overcome, the internal kinetics guide the state vector directly into the divergence free modes dictated by the network geometry and observed in experiments \cite{arava2025energy}.
\section{\label{large}Large open clusters}To study non-equilibrium relaxation in open spin networks, we use a half-Husimi cactus representation, a connected network constructed by sharing closed square plaquettes. This structure preserves local square-plaquette geometry ($C_4$ cycles), vertex coordination ($z=1,2,3,4$), and ice-rule frustration while eliminating long-range closed back-edges.
Rather than a homogeneous assembly, the finite half Husimi cactus \cite{otsuka2018husimi,udagawa2019spectrum,udagawa2016out} forms a structurally non uniform spatial environment for magnetic flux relaxation. It consists of a semi-infinite Husimi tree rooted at a single open boundary link ($n=0$) and branching radially into $g$ generation shells. Its radially varying coordination numbers prevent a magnetized state from remaining charge neutral everywhere. While the interior bulk junctions with coordination $z=4$ remain charge compensated, the applied magnetizing field forces the truncated boundary leaves (outer peripheral boundary nodes) with $z=2$ into charged configurations. The cluster thus starts as a stressed network with peripheral monopole reservoirs surrounding charge compensated interior valleys. As a result, global relaxation combines the kinetic regimes of the Loop and the Vertex. The $z=2$ boundaries resemble the charged monopole loop: with high initial monopole chemical potential, their charges rapidly annihilate or flow ballistically inward without an initiation barrier. In contrast, the $z=4$ bulk behaves like an isolated neutral vertex, undergoing strong kinetic arrest and relaxing only via stochastic, thermally activated creation of transient monopole pairs. The global composite operator, a non homogeneous transfer matrix chain $\tilde{\mathcal{M}} = \mathcal{P}_{\text{cl}} \prod_{n} \mathbf{T}_{n \leftarrow n-1}$, couples these environments. This intersection of transport regimes is expected to produce the intermittent, multiscale, non Arrhenius kinetics and bimodal avalanche distributions.
\\
In out of equilibrium systems, relaxation is often tracked by thermodynamic order parameters such as net magnetization  \cite{mellado2010dynamics}. Because transport on the half Husimi cactus is non homogeneous along its tree like hierarchy, we instead use a probabilistic transport metric. The total energy of any vertex configuration is given by Eq.~\ref{eq1}. As the system relaxes from its initial state, a single spin flip advances the configuration space from generation $n-1$ to $n$. The local activation energy barrier $\Delta E_n$ is the energy difference between the transient state (saddle point) and the initial configuration, and represents the total activation energy barrier (which includes the local vertex self-energies plus the incremental nearest-neighbor Coulomb interaction term). Boundary zones (Scenario I) give $\Delta E_0 \sim 0$ because boundary charges dissipate inward. In contrast, deep interior junctions ($n>1$, Scenario II) incur a large cost, $\Delta E_n = 4\chi + \delta E_{\text{Coulomb}}$, the energy to create a localized monopole–antimonopole pair from a compensated center ($Q_v=0$). The matrix elements of the layer transfer operator $\mathbf{T}_{n \leftarrow n-1}$ scale with temperature via an Arrhenius factor,
\begin{equation}
   \mathcal{R}_n(T) = \exp\left(-\frac{\Delta E_n}{k_B T}\right)
\end{equation}
The initial kinetic arrest probability of a layer is the ratio of its trapped configuration weight to its total local path partition function. Because the half Husimi cactus is a branching architecture, it introduces a   path degeneracy $\omega_n$ by counting the equivalent geometric routes or fundamental cycles available for flux propagation in that generation. The local initial arrest probability at generation $n$ reduces to
\begin{equation}
    \mathcal{P}_n(T) = \frac{1}{1 + \omega_n [\mathcal{R}_n(T) ]} 
\end{equation}
Because the global composite transfer matrix $\mathcal{M}$ is a linear chain of non homogeneous layer operators, complete initial arrest requires every layer to remain blocked. Requiring that no layer undergoes a thermal defect creation event over a short observation window, the global initial arrest probability $\mathcal{P}(T)$ of a half Husimi cactus of $g$ generations is the product of the blocking probabilities of all individual layers:
\begin{equation}
    \mathcal{P}(T,g) = \prod_{n=0}^{g} \frac{1}{1 + \omega_n [\mathcal{R}_n(T)]}.
    \label{eqpgh}
\end{equation}
Throughout, ``initial arrest'' refers to the survival of the prepared state over this window ($t\ll\tau_{\rm escape}$), not to a long-time property.

For example, in a 3 generation half Husimi cactus:
\begin{itemize}
\item $n=0$: The root vertex ($z=1$) where the unconstrained dangling link anchors the cluster. Its   degeneracy is $\omega_0 = 1$ (one open input link), and its activation energy is $\Delta E_0 = 0$. As a dead end boundary edge, flipping this spin costs no local self energy. The blocking probability is $\mathcal{P}_0 = \frac{1}{1 + 1} = \frac{1}{2}$.
\item $n=1$: The central base junction ($z=3$) where the dangling link ends (central trunk connecting the root to the first square). This vertex branches symmetrically into left and right paths, forming the base of the first central square plaquette with $\omega_1 = 2$. Changing its state costs a vertex self energy plus near neighbor Coulomb corrections, $\Delta E_1 = \chi + \delta E_{\text{Coulomb}}^{(1)}$. The blocking probability is
\[
\mathcal{P}_1(T) = \frac{1}{1 + 2 \exp\left(-\frac{\chi + \delta E_{\text{Coulomb}}^{(1)}}{k_B T}\right)},
\]
where 
\[
\delta E_{\text{Coulomb}}^{(1)} = \sum_{v \in \text{nn}(1)} \frac{Q_{(1)} Q_v}{r_{1,v}}
\]
denotes the dimensionless Coulomb interaction energy arising from nearest-neighbor coupling between the boundary vertex charge $Q_{(1)}$ and the neighboring charges $Q_v$ in generation 2. In this expression, $r_{1,v}$ is the spatial distance separating vertex 1 from node $v$, with the sum taken over the immediate nearest-neighbor set $\text{nn}(1)$.
\item $n=2$: Moving along the paths of the first central square, we encounter the internal bulk nodes ($z=4$), fully coordinated bulk vertices where the central square branches into higher generations. In a symmetric tree, each of the two $n=1$ paths meets a $z=4$ vertex that splits into two new paths, giving $2 \times 2 = 4$ paths through this shell and a   degeneracy $\omega_2 = 4$. These deeply embedded, initially charge neutral nodes ($Q_v = 0$) can only be reconfigured via Scenario II, i.e., creation governed dynamics. Driving a domain wall across this layer requires the spontaneous creation of a monopole–antimonopole pair, incurring the maximum energy cost $\Delta E_2 = 4\chi + \delta E_{\text{Coulomb}}^{(2)}$. The blocking probability is 
\[
\mathcal{P}_2(T) = \frac{1}{1 + 4 \exp\left(-\frac{4\chi + \delta E_{\text{Coulomb}}^{(2)}}{k_B T}\right)}.
\]
\item $n=3$: Truncated leaves ($z=2$). The vertex marked $z=2$ is the shared corner where the system ends at the boundary. The degeneracy $\omega_3$ captures the large path multiplicity of the outer boundary loops. $\Delta E_3 \sim 0$ because these outermost $z=2$ leaves already host uncompensated magnetic charges in a macroscopically magnetized initial state, giving them a high initial monopole chemical potential. They follow Scenario I, where charges glide inward or annihilate immediately with no creation barrier, $\mathcal{P}_3 = \frac{1}{1 + \omega_3}$.
\end{itemize}
Overall, the global initial arrest probability of a 3 generation half Husimi cactus becomes
\begin{widetext}
\begin{equation}
\mathcal{P}(T,3) = \frac{1}{2(1 + \omega_3)} \left[ \frac{1}{1 + 2 \exp\left(-\frac{\chi + \delta E_{\text{Coulomb}}^{(1)}}{k_B T}\right)} \right] \left[ \frac{1}{1 + 4 \exp\left(-\frac{4\chi + \delta E_{\text{Coulomb}}^{(2)}}{k_B T}\right)} \right]
\end{equation}
\end{widetext}
The constant boundary reduction factors ($\frac{1}{2}$ at the root and $\frac{1}{1+\omega_3}$ at the leaves) are purely geometric; the thermal gating of the network is controlled exclusively by the interior bulk junctions ($z=3$ and $z=4$). As temperature drops, the creation barrier at the $z=4$ bulk sites suppresses the third term to $1$, locking the system into a global initial kinetic arrest. This demonstrates how structural coordination numbers translate directly into a hierarchical cascade of sequential energy barriers. For an arbitrary generation depth, the global initial arrest probability is obtained by tracking the geometric scaling of the path degeneracies and the coordination dependent activation barriers across the hierarchy. Radially, the path multiplicity scales exponentially within the bulk as $\omega_n = 2^n$, while the boundaries are anchored by $\omega_0 = 1$ at the dangling root.
\\
To isolate the intrinsic thermodynamic behavior of the hierarchical interior, separate from the static boundary leaf counts ($\frac{1}{2(1 + 2^g)}$), we factor out temperature-independent geometric prefactors and introduce the interior bulk blocking probability $\mathcal{P}_{\text{bulk}}(T,g)$ over successive generations $1 \leq n < g$: \begin{equation} \mathcal{P}_{\text{bulk}}(T,g) = \prod_{n=1}^{g-1} \frac{1}{1 + 2^n \exp\left( -\frac{\Delta E_n}{k_B T} \right)}. \label{eq:p_bulk_normalized} 
\end{equation} 
where the layer dependent activation energy $\Delta E_n$ transitions from a single vertex type penalty at the base junction to a full defect creation barrier within the core
\begin{equation}
    \Delta E_n = \begin{cases} 
    \chi + \delta E_{\text{Coulomb}}^{(1)} & \text{for } n = 1, \\ 
    4\chi + \delta E_{\text{Coulomb}}^{(n)} & \text{for } 1 < n < g. 
    \end{cases}
\end{equation}
with $\delta E^{(n)}_{\rm Coulomb} =-|Q_uQ_w|/r=-4$. For all interior bulk layers, the defect-creation activation barriers are strictly positive ($\Delta E_n > 0$, $\chi>1$). In the low-temperature limit ($T \to 0$), thermal excitation of monopole pairs freezes out, causing the exponential Boltzmann factors to vanish ($\exp\left(-\frac{\Delta E_n}{k_B T}\right) \to 0$). The product denominators reduce to unity, yielding: \begin{equation} \lim_{T \to 0} \mathcal{P}_{\text{bulk}}(T,g) = \prod_{n=1}^{g-1} (1) = 1. \end{equation} This confirms that at absolute zero, the system saturates at complete initial kinetic arrest ($\mathcal{P}_{\text{bulk}} = 1$), in agreement with Fig.~\ref{f3}a. Conversely, in the high-temperature limit ($T \to \infty$), thermal energy overcomes all activation barriers, and the exponential terms approach unity, dropping the bulk blocking probability to its minimum asymptotic floor: \begin{equation} \lim_{T \to \infty} \mathcal{P}_{\text{bulk}}(T,g) = \prod_{n=1}^{g-1} \frac{1}{1 + 2^n}. \end{equation} As the generation depth $g$ grows, this product rapidly vanishes ($\mathcal{P}_{\text{bulk}} \to 0$), representing a completely fluid, barrier-free transport regime.
\\
Since $\mathcal{P}_{\text{bulk}}(T, g)$ is composed of several multiplicative terms, each generation layer effectively functions as an independent thermal gate that turns on at its own characteristic temperature scale, $T_c(n) \sim \frac{\Delta E_n}{k_B \ln(2^n)} = \frac{\Delta E_n}{n k_B \ln(2)}$. Because $\Delta E_{\rm{bulk}}$ is identical for all interior layers, the scales $T_c(n) = \Delta E_{\rm{bulk}}/(k_B n \ln{2})$ are split by the path multiplicity $\omega_n = 2^n$ alone, and deeper layers freeze at lower T. For a fixed generation depth $g$, $\mathcal{P}_{\text{bulk}}(T,g)$ represents a smooth, multi-step cascade profile constrained by the network’s geometry. Limiting magnetic Coulomb interactions to nearest neighbors simplifies the energy landscape, as each vertex charge interacts only with adjacent charges at the nearest neighbor separation r=1. Since the starting magnetized configuration places uncompensated charges only on the outer boundary, vertices well inside the bulk feel no baseline Coulomb potential from the perimeter and incur electrostatic costs only from their immediate neighbors in adjacent layers. At the base junction ($n=1$), changing the vertex state costs a single vertex self-energy penalty plus nearest-neighbor Coulomb corrections, $\Delta E_1 = \chi + \delta E_{\text{Coulomb}}^{(1)}$. In the interior bulk region ($1 < n < g$, $z=4$), transmitting an excitation requires creating a nearby monopole–antimonopole pair on adjacent nodes. The associated energy cost is given by the self-energy ($4\chi$) for the two $Q=\pm 2$ vertices of the pair, offset by the attractive nearest-neighbor Coulomb binding energy of the newly created pair
\begin{equation} 
\begin{aligned} 
\Delta E_n &= 4\chi - \frac{|Q_u Q_w|}{r} = 4(\chi - 1) \equiv \Delta E_{\text{bulk}} \\ &\quad \text{for } Q=\pm 2, \; r=1, \; \chi>1
\end{aligned} \end{equation}
Because $\Delta E_{\rm{bulk}}$ is identical for every interior layer, independent of the layer index $n$ and overall system depth $g$, the core of the global product reduces to a homogeneous progression. With $k = e^{-\beta \Delta E_{\rm bulk}}$, the global interior bulk blocking probability simplifies to: \begin{equation} 
\mathcal{P}_{\text{bulk}}(T, g) = \left[ \frac{1}{1 + 2 \exp\left(-\frac{\chi + \delta E_{\text{Coulomb}}^{(1)}}{k_B T}\right)} \right] \prod_{n=2}^{g-1} \frac{1}{1 + 2^n k}. \label{eqPG} 
\end{equation}

Because trees are inherently recursive, we can isolate the size dependence by defining $\mathcal{K}(g, T) = \prod_{n=2}^{g-1} \frac{1}{1 + 2^n k}$. This product satisfies a simple recursive relation when shifting from system size $g$ to $g+1$: \begin{equation} \mathcal{K}(g+1, T) = \mathcal{K}(g, T) \frac{1}{1 + 2^g k}. \end{equation}
In the scaling window $T\sim x^*/g$ the factor of layer $n$ differs from one by
$O(2^{\,n-g\Delta E_n/\Delta E_{\rm bulk}})$. Fig.~3 is evaluated with
$\Delta E_n=\Delta E_{\rm bulk}$ for all $1\le n<g$; for a lower first-layer barrier
$\Delta E_1<\Delta E_{\rm bulk}$ the same collapse is reached for
$g\gg\Delta E_{\rm bulk}/(\Delta E_1\ln2)$.
\subsection{Experimental validation on Small Finite Clusters}
To qualitatively validate the transfer-matrix barrier predictions against real physical systems, we compare our theoretical energy trajectories against Magnetic Force Microscopy (MFM) and AC demagnetization data on artificial square-ice sub-clusters \cite{arava2025energy}. The relaxation pathways fall into two distinct regimes dictated by local vertex coordination and dipolar energy landscapes:

i) \textit{Monotonic/Non-Arrested Geometries (Loop-like, Fish, H, and Triple-Vertex Clusters):} In these geometries, intermediate defect states are sequentially lower or equal in dipolar magnetostatic energy. Because relaxation proceeds along monotonic downhill energy trajectories without encountering high-energy vertex creation penalties ($z \le 3$), monopole pairs annihilate smoothly, yielding high experimental ground-state fractions of \text{72\%--82\%}.

ii) \textit{Intermittent/Kinetically Arrested Geometries (Vertex-like, Splitfish, Toffee, and Antenna Clusters):} In configurations containing central $z=4$ junctions, relaxation requires passing through high-energy intermediate states that carry $\pm 2q$ monopole charges. These states introduce significant dipolar energy hills that act as bottlenecks, trapping the system in metastable configurations and causing bimodal pathway splitting with reduced ground-state yields of \text{47\%--52\%}.

The agreement between local vertex coordination and experimental relaxation points to the predictive power of the transfer-matrix formalism. Comparing the calculated dipolar energy trajectories $\Delta U(\boldsymbol{\sigma})$ across relaxation steps reveals a fundamental distinction between monotonic and intermittent cluster families \cite{arava2025energy, arava2026geometry}. For monotonic geometries, the transfer operator spectrum is dominated by a single fast decay mode ($\tau_{\text{fast}} \sim \nu_0^{-1}$), reflecting coordination-unconstrained charge removal at open boundaries. Conversely, for intermittent geometries containing $z=4$ core vertices, the spectrum exhibits a large spectral gap separating $\tau_{\text{fast}}$ from a slow bulk creation mode ($\tau_{\text{slow}} \sim \nu_0^{-1} e^{\Delta E_{\rm bulk} / k_B T}$). This barrier structure qualitatively accounts for the bimodal yield split (72\%--82\% vs.\ 47\%-52\%) observed in MFM demagnetization experiments \cite{arava2025energy, arava2026geometry}.

\section{\label{finite}Finite size scaling}
Using the local approximation, we have carried out a finite size scaling analysis on the global initial arrest probability, where the number of generation layers $g$ plays the role of the characteristic linear system size. As shown in Fig.~\ref{f3}a, increasing the generation depth causes the product of the non homogeneous layer operators to steepen, asymptotically sharpening the global blocking probability from a smooth crossover into a discrete, step like discontinuity at a pseudo-critical temperature $T_c=x^*/g\to 0$.
\\
At temperature $T$ the $n$-th layer is thermally active when $\omega_nR_n\gtrsim1$, i.e.\ for
$n\gtrsim\xi(T)$, while layers $n<\xi$ remain blocked. We take $\xi$, measured in
generations, as the crossover length,
\begin{equation}
\begin{aligned}
\xi(T)=\frac{\Delta E_{\rm bulk}}{k_BT\ln2}=\frac{x^*}{T}\sim T^{-\nu},\quad \nu=1,
\\ x^*=\frac{4(\chi-1)}{\ln2} 
\end{aligned}
\end{equation}
Since $\omega_nR_n=2^{\,n-\xi}$, relabeling $m=g-n$ in Eq.~(22) gives
$P_{\rm bulk}=\prod_{m=1}^{g-1}[1+2^{\,z-m}]^{-1}$, i.e.
\begin{equation}
\begin{aligned}
P_{\rm bulk}(T,g)\approx F\big(g-\xi(T)\big)=\mathcal{F}\!\left[g\Big(1-\frac{x^*}{gT}\Big)\right],\\
\mathcal{F}(z)=\prod_{m\ge1}\frac{1}{1+2^{\,z-m}}. 
\end{aligned}
\end{equation}
where $z\equiv g-\xi(T)$ and $\mathcal{F}$ is a dimensionless scaling function dictated entirely by the structural constraints of the network \cite{barber1983finite}.

We collapsed the calculated $P_{\rm bulk}(T,g)$ profiles for $g=10$--$50$.
Plotted against $x=gT$ (Fig.~3b), the curves cross at $x^*=4(\chi-1)/\ln2\simeq1.44$, the
point where the path multiplicity balances the activation barrier $\Delta E_{\rm bulk}$.
Plotted against $z=g-x^*/T$ (Fig.~3c) they collapse onto the single function $\mathcal{F}(z)$ of
Eq.~(30), establishing $\nu=1$. Since $T_c\to0$, this is a finite-size scaling crossover
associated with kinetic arrest, not a finite-temperature transition. The coordination-driven
kinetic arrest is reminiscent of athermal granular jamming, but it stems from the spatial
coordination landscape of finite magnetic networks rather than from mechanical contact
constraints \cite{liu1998jamming,pica2009jamming}.
\section{\label{conclusions}Conclusions and Outlook}
The implementation of a non homogeneous transfer matrix chain on a half Husimi cactus provides an algebraic foundation for characterizing the evolution of finite magnetic networks. By mapping the non uniform coordination hierarchy, the model accounts for the uncompensated moments at boundary extremities that drive relaxation kinetics. This formalism is consistent with dynamic observations, such as kinetic arrest and telegraph noise, since it derives transition probabilities from localized Boltzmann weights. 
\\
By quantifying the probability of initial kinetic arrest at specific nodes, the model distinguishes between self propelled monopole annihilation in boundaries and the   locking of compensated boundaries. Thus, at the level of nearest neighbor Coulomb interactions, it provides a map of flux propagation, identifying precisely which coordination zones act as the dominant physical bottlenecks in magnetic evolution. The principal eigenvector of a composite operator constructed from a sequence of transfer matrices is dominated by charge-neutral configurations, reflecting the free-energy advantage of the ice-rule manifold. The chiral antiferromagnetic ground state is resolved in the Loop building block, and its selection in larger clusters is consistent with the charge-level analysis. Simultaneously, the relaxation times of the Markov generator are expected to be controlled at low temperatures by the bottleneck barrier.
\\
The relaxation processes examined in this work describe the out of equilibrium dynamics of time irreversible relaxation pathways in finite magnetic networks initialized in a magnetized state. Rather than focusing on equilibrium fluctuations, we study the propagation of flux through a coordination hierarchy, where a strong separation of timescales is expected to give rise to non equilibrium phenomena, including kinetic arrest and rapid, avalanche like relaxation of configurations.
\\
The hierarchical transport obtained on the half Husimi cactus offers a direct counterpart to the out of equilibrium relaxation observed in finite, closed artificial spin ice square arrays. In such systems, large scale relaxation is dictated primarily by boundary conditions rather than by bulk transport parameters. Peripheral vertices with low coordination and open roots function as   monopole sources and sinks, injecting magnetic charges into the network at a low activation cost. As these excitations propagate inward, they depart from the unconstrained boundaries and enter the densely coordinated interior bulk, where they are trapped in charge neutral, creation dominated potential wells. Because the system has a finite size, the spectrum is discrete, so relaxation appears as a hierarchy of time scales. As a result, relaxation appears as a distinctly quantized hierarchy of time scales, spanning from fast, ballistic annihilation at the boundaries to intermittent avalanches in the central region. Applying a finite-size scaling analysis, we show that the truncation of the crossover length $\xi(T)$ by the cluster depth g produces the size-dependent crossover of the initial arrest probability, demonstrating that finite boundaries alone can generate jamming-like arrest in the absence of extrinsic disorder.
\\
Our transfer-matrix framework estimates activation barriers from local vertex self-energies and nearest-neighbor Coulomb terms. Long-range dipolar and higher-order multipolar screening mainly renormalize the barriers’ numerical magnitudes rather than changing the fundamental physical picture. Since nonlocal magnetostatic interactions fall off quickly with distance, long-range screening produces a smooth drift in local activation thresholds but preserves the discrete energy increments set by the coordination hierarchy ($z = 1, 2, 3, 4$). We therefore anticipate that the coordination bottlenecks underlying boundary gating, the bimodal relaxation time scales, and the finite-size crossover will remain qualitatively robust even when long-range magnetostatic corrections are included. Incorporating a truncated or screened multipolar potential into the transfer-operator chain would help elucidate how long-range interactions modify the spatial cascade structure of the localized energy barriers identified here.
\\
As a concluding note, we point out that drawing a direct link between coordination zones and kinetic bottlenecks provides a basis for “geometrical gating”. By lithographically tuning boundary geometries, vertex types, and the generation depth of finite networks, it becomes possible to precisely steer the direction, rate, and quantized nature of flux propagation. This creates prospects for engineered nanomagnetic logic gates, neuromorphic computing architectures, and multistate magnetic memory elements governed purely by geometric design.
\\
The framework can be extended to Husimi trees built from triangles or hexagons, which mimic kagome and honeycomb spin ices. Because $\nu= 1$ follows from the exponential growth of the path multiplicity, we expect the collapse in $g-\xi(T)$ to persist, with a geometry dependent crossing point $x^*$ set by the branching ratio and the activation barrier. Testing this would establish how general flux jamming is across coordination hierarchies.

\begin{acknowledgments}
P.M. acknowledges support from the Fondo Nacional de Desarrollo Científico y Tecnológico (Fondecyt) under Grant No. 1250122. H.A. was funded by the US Department of Energy, Office of Science, Office
of Basic Energy Sciences, Materials Science and Engineering Division. 
\end{acknowledgments}
%\bibliography{graph}

\end{document}